\documentstyle[11pt,aaspp]{article}

\newcommand{\epscma}{$\epsilon$ CMa}
\newcommand{\betcma}{$\beta$ CMa}    
\newcommand{\etauma}{$\eta$ UMa}
\newcommand{\spica}{$\alpha$ Vir}
\newcommand{\kms}{\hbox{$\,$km$\,$s$^{-1}$}}
\newcommand{\Teff}{\hbox{$\,T_{\rm eff}$} }

\newcommand{\Logg}{\hbox{$\log g$} }

\newcommand{\AD}{\hbox{$\,\theta_d$} }
\newcommand{\phoenix}{{\tt PHOENIX}}
\newcommand{\oaotwo}{{\it OAO-2} }
\newcommand{\iue}{{\it IUE} }

\newcommand{\tdone}{{\it TD-1} }
\newcommand{\ghrs}{{\it GHRS} }

\lefthead{Aufdenberg et al.}
\righthead{NLTE Model of \epscma}

\begin{document}

\title{A Spherical Non-LTE Line-Blanketed Stellar Atmosphere Model of 
the Early B Giant \epscma}

\author{J. P. Aufdenberg\altaffilmark{1}}
\affil{Department of Physics and Astronomy, Arizona State University,
    Tempe, AZ 85271-1504}

\author{P. H. Hauschildt\altaffilmark{2}} 
\affil{Department of Physics and Astronomy and Center for Simulational 
Physics,\\ University of Georgia, Athens, GA 30602-2451}

\author{S. N. Shore\altaffilmark{3}} 
\affil{Department of Physics and Astronomy, Indiana University South Bend, 
South Bend, IN 46634-7111}

\author{E. Baron\altaffilmark{4}}
\affil{Department of Physics and Astronomy, University of Oklahoma,
Norman, OK 73019-0225} 

\altaffiltext{1}{Email: jaufdenb@sara.la.asu.edu} 
\altaffiltext{2}{Email: yeti@hal.physast.uga.edu}
\altaffiltext{3}{Email: sshore@paladin.iusb.edu}
\altaffiltext{4}{Email: baron@mail.nhn.ou.edu}

\begin{center}
{\bf Accepted for Publication in the Astrophysical Journal\\}
{\bf To Appear in the 10 May 1998 Issue}\\
{\bf also available at \tt ftp://calvin.physast.uga.edu/pub/preprints/epsCMa.ps.gz}

\end{center}

\begin{abstract}

We use a spherical non-LTE fully line blanketed model atmosphere to
fit the full multi-wavelength spectrum, including the extreme
ultraviolet (EUV) continuum observed by the {\it Extreme Ultraviolet
Explorer}, of the B2 II star \epscma.  The available spectrophotometry
of \epscma\ from 350 \AA\ to 25 \micron\ is best fit with model parameters
\Teff = 21750\,K, \Logg = 3.5, and an angular diameter of 0.77 mas.  
Our best fit model predicts a hydrogen ionizing flux, $q_0$, of
$1.59\times10^{21}$ photons cm$^{-2}$ s$^{-1}$ at the star's surface
and 2290 photons cm$^{-2}$ s$^{-1}$ at the surface of the Local Cloud.

The close agreement between the model and the measured EUV flux from
\epscma\ is a result of the higher temperatures at the formation
depths of the \ion{H}{1} and \ion{He}{1} Lyman continua compared to
other models.  The realistic model treatment of early B giants with 
spherical geometry and NLTE metal line blanketing results in the
prediction of significantly larger EUV fluxes compared with
plane-parallel models.  We find that our metal line blanketed
spherical models show significantly warmer temperature structures, 1-3
kK at the formation depth of the Lyman continua, and predict stronger
EUV fluxes, up to a factor of 5 in the \ion{H}{1} Lyman continuum,
compared with plane-parallel atmospheres that have identical model
parameters.  In contrast, we find spherical and plane-parallel models
that do not include metal line blanketing are nearly identical.  Our
\Teff = 21000 K, \Logg = 3.2, spherical NLTE model predicts more than
twice as many hydrogen ionizing photons and over 200 times more
neutral helium ionizing photons than a standard hydrostatic
plane-parallel LTE model with the same stellar parameters.

Our synthetic spectra are in reasonably good agreement with observed
continuum and line fluxes from echelle spectra obtained with the
Goddard High Resolution Spectrograph.  While we find agreement between
the absolute UV flux of \epscma\ as measured by {\it GHRS} and our
model atmosphere, these fluxes are $\sim$30\% higher in the UV than
those measured by {\it IUE}, {\it OAO-2}, and {\it TD-1}, in excess of
the published errors in the absolute calibration of these data.

\end{abstract}


\section{Introduction}

Spectroscopy of the bright B2 II star \epscma\ (HD 52089, HR 2618)
(\cite{epscma95}) below the Lyman edge by the {\it Extreme Ultraviolet
Explorer (EUVE)} satellite is possible due to the extremely low
neutral hydrogen column density toward this star.  As a result of its
location at a {\it HIPPARCOS} distance of 132$^{+11}_{-9}$ pc along an
exceptionally rarefied interstellar tunnel extending out from the
Local Bubble (\cite{welsh91}), \epscma\ is attenuated by a neutral
hydrogen column density of less than $5\times 10^{17}\ {\rm cm^{-2}}$
(\cite{gry95}).  Consequently, \epscma\ is an extremely important star
for its contribution to the local interstellar hydrogen ionization,
producing more hydrogen ionizing flux than all nearby stars combined
(\cite{vallerga}).  The large EUV flux from \epscma\ affects the
ionization state of the Local Cloud, the region of neutral hydrogen
concentrated within a few parsecs of the Sun, in which the solar
system is embedded (\cite{LISM}).

The observation of the stellar continuum below the Lyman edge provides
an opportunity to directly compare the observed flux distribution of a
hot star with the extreme ultraviolet (EUV) flux predicted by model
stellar atmospheres.  Model atmospheres of hot stars are generally in
good agreement with the observed fluxes longward of 912 \AA.  The lack
of spectroscopic data in the extreme ultraviolet has, until very
recently, left models untested in the EUV.  Thus, the observations
that the continua of \epscma\ and \betcma\ observed by {\it EUVE}
exceed the fluxes predicted by the present model atmospheres, by a factor
of 5 in the case of \betcma\ (\cite{betacma95}) and by more than an order
of magnitude in the case of \epscma\ (\cite{epscma95,naj96}), 
pose a serious challenge to such computations.

These stars exhibit an ``EUV excess'' relative to the predictions of
the hydrostatic plane-parallel line blanketed LTE models of Kurucz
(1992) and the non-LTE models of
\cite{hublanz95}.  The non-LTE spherically extended atmosphere models
of Najarro et al. (1996), that include stellar winds, but neglect metal
line blanketing, also do not correctly predict the observed EUV flux.  The
non-LTE atmosphere models of Schaerer \& de~Koter (1997), which include stellar
winds and line blanketing predict the EUV flux distribution of 
\epscma\ and \betcma, but neglect line broadening, 
which artificially restricts the blanketing effects.
When line broadening is included, Schaerer \& de~Koter find a significant
decrease in the predicted EUV flux.  These models fail to predict the
EUV flux presumably because they do not correctly predict either the
temperature structure or level populations of H and \ion{He}{1} in
the continuum forming layers.  It has also been suggested that the temperature
structure and level populations are affected by the presence of the
stellar wind (\cite{naj96,schaerer3}) and by X-ray heating (\cite{cohen96}).

We present here hydrostatic, spherical, non-LTE, metal line blanketed
model atmospheres and synthetic spectra which closely reproduce both
the observed EUV continuum and the flux distribution longward of 912
\AA\ to 25 \micron\ of \epscma.  In section 2 we describe the
calculation of our model atmospheres.  In section 3 we compare the
synthetic spectra to observational data.  We discuss our results and
conclusions in section 4.

\section{Stellar Atmosphere Models}

We have computed stellar atmosphere models with the generalized
stellar atmosphere computer code \phoenix\ (version 8.1). It
calculates LTE and non-LTE, line blanketed, spherical expanding model
atmospheres (\cite{novaphys,fe2nova,parapap,parapap2} and references
therein).  Although \phoenix\ is able to treat stellar winds, we do
not consider the effect of a stellar wind here.  In addition, we do
not consider here the effects of rotation, X-ray heating or magnetic
fields. \phoenix\ can handle very large non-LTE model atoms as well as
line blanketing by millions of atomic (and molecular) lines.  This
code is designed to be very flexible.  It has been used to compute
model atmospheres and synthetic spectra for novae, supernovae, white,
brown, and M dwarfs, and accretion disks in Active Galactic Nuclei.
The code is very well tested for plane-parallel and static cool stars
(\cite{MDpap}).  In the following paragraphs we give a brief
description of its most salient features; more detailed descriptions
can be found in the references.

We use here an equation of state including up to 26 ionization stages
of 39 elements. The large number of ionization stages is required to
span the wide range of electron temperatures and gas pressures
encountered in OB-star atmospheres.
The models also include all relevant bound-free and
free-free processes with the cross sections given in the compilations
of \cite{mathisen} and \cite{verner95}.  The low densities and
complicated radiation fields in the atmospheres require that the most
important species be treated with self-consistent multi-level NLTE
using the methods described in \cite{casspap}, \cite{fe2pap}, and
Hauschildt, Baron, \& Allard (1997).  Line blanketing
is a  fundamental effect in the computation of OB star model
atmospheres.  Therefore, we include, in addition to the NLTE primary
and secondary lines, the most important lines
dynamically selected from the list of $42\times 10^{6}$ lines of
\cite{cdrom1}; for details of the selection process and the treatment
of these lines see Hauschildt et~al. (1995), Allard \& Hauschildt
(1995) and references therein.  All models assume solar abundances 
(\cite{andgrev89}).

In order to allow for line scattering in LTE background lines, which
is important in low density plasmas, we parameterize the albedo for
line scattering using a parameter $\alpha$ (\cite{novapap}).  A
fraction, $\alpha$, of the line absorption coefficient is assumed to
contribute to the absorption coefficient while the remaining fraction
(1-$\alpha$) is included as part of the line scattering.  This is done
only for lines of species {\em not} treated explicitly in NLTE.  For
our NLTE models we use $\alpha$ = 0.95.  In our LTE models we use
either $\alpha$ = 0.0 for no line scattering, pure LTE, or we use
$\alpha$ = 0.95 to include line scattering.  The effects of LTE
line scattering on the temperature structure and synthetic spectra are
discussed in section 4.
  
We include the effects of statistical (random) velocity fields
using a Gaussian broadening velocity $\xi$. For the models presented
here, we use $\xi=2 \kms$.  Each line is calculated with its
individual depth dependent intrinsic profile.  For strong lines we use
Voigt profiles, while for weak lines we use Gaussian profiles (\cite{vb10pap}).

The standard outer boundary conditions for our model atmospheres are a
continuum optical depth of $\tau_{\rm 2000\AA}$ = $10^{-10}$ at 2000
\AA\ and an outer gas pressure of P$_{\rm gas}$ = $10^{-4}$ dynes
cm$^{-2}$. The inner
optical depth boundary is $\tau_{\rm 2000\AA}$ = $10^{2}$.  
Models are computed on an optical depth grid of 50 points. 
The effects of these boundary conditions and other model parameters on the
temperature structure and synthetic spectrum are discussed in detail in
section 4.

Our solution of the radiative transfer equation is based on an
operator splitting method with an exact band-matrix form of the
approximate $\Lambda$-operator, see \cite{s3pap} and \cite{aliperf}
for details.  Our numerical solution of the radiative energy equation
is based on an improved version of the Uns\"old-Lucy method as
described in Allard \& Hauschildt (1995).  The temperature structure
is computed from the condition of radiative equilibrium.  The
temperature corrections are based on the absolute error in both the
flux and the flux derivative.  Models are typically converged to
better than 1.5\% in the absolute flux and the flux derivative at every 
depth. The maximum relative change in the temperature at any depth is not
greater than 0.8\% and better than 0.2\% in the EUV continuum forming
layers.

We have computed four different classes of models:  LTE models, NLTE
models with H and He but without metal line blanketing, NLTE models
including ions of H, He, C, N, O, and NLTE models including ions of H,
He, C, N, O, Ca, Mg, and Fe. These classes of models are
described below.  Additional detailed information on the NLTE model
atoms can be found in Hauschildt et~al. (1995), Hauschildt
et~al. (1996), Hauschildt, Baron \& Allard (1997), Hauschildt et~al.
(1997), and references therein. The models discussed in this paper are
listed in Table 1.

1) Typical LTE models (and NLTE models which consider background line
blanketing in LTE) use lines from ions of H, He, Li, Be, B, C, N, O,
F, Ne, Na, Mg, Al, Si, P, S, Cl, Ar, Ca, Sc, Ti, V, Cr, Mn, Fe, Co,
Ni, Cu, Zn, Ga, Kr, Rb, Sr, Y, Zr, Nb, Ba, and La selected from
\cite{cdrom1}.  LTE lines are selected for inclusion in the model
atmosphere based on the strength of the line absorption coefficient at
line center, $\kappa_{\rm lc}$, relative to the absorption coefficient
of surrounding continuum, $\kappa_{\rm cont}$.  For most of the models
considered here, selected LTE lines satisfy $\kappa_{\rm
lc}/\kappa_{\rm cont} > 10^{-4}$.  Under this condition, approximately
$2.1\times10^6$ lines are considered, of which most are treated as
Gaussian profiles, while the $9.4\times10^3$ strongest lines
($\kappa_{\rm lc}/\kappa_{\rm cont} > 10^{2}$) are calculated with
Voigt profiles.  In section 4 we discuss the effects of the number of
blanketing lines on the model solution.  The model is computed on a
grid of 11074 wavelength points.  The LTE models converged using
higher resolution grids are discussed in section 4.  The LTE models
are calculated starting from an LTE-grey model and iterated to
convergence in 30 iterations, 767 CPU seconds per iteration, on a HP
9000/770 workstation.

2) An NLTE pure H-He model with no metal line blanketing uses 101
levels, producing 545 transitions from the ions H I and He I-II.  The
models are computed on a grid of 13569 wavelength points. The models
are started from an LTE-grey model structure and iterated to
convergence in 30 iterations, 782 CPU seconds per iteration, on an HP
9000/770 workstation.

3) A typical NLTE H-He-CNO model includes 1751 distinct levels, producing
11975 transitions from the ions H I, He I-II, C I-IV, N I-VI, and O
I-VI in detailed NLTE in addition to the background LTE line
blanketing discussed above.  The models are computed on a grid of
63486 wavelength points.  The NLTE models are calculated starting from
a converged LTE line blanketed model and iterated to convergence in 30
iterations, 7707 CPU seconds per iteration, on an HP 9000/770
workstation.

4) More complete NLTE models employ 3035 levels, with 37151 transitions
from the ions H I, He I-II, C I-IV, N I-VI, O I-VI, Ca II, Mg II, and
Fe II-III in  NLTE in addition to the LTE line blanketing.
These models are computed on a wavelength grid of 183591 wavelength
points.  The models are calculated starting from a NLTE
H-He-CNO class model and converged in 10 iterations, 8900 CPU seconds
per iteration, running the parallel version of the \phoenix\ code on an
IBM SP2 computer at the Cornell Theory Center.

%% FIGURE 1
\begin{figure}
\caption{The continuous energy distribution of \epscma\ from 350 \AA\ to
25 \micron\ compared with the 21750 K NLTE model (N1) 
and the ATLAS9 21000 K model.  The {\it EUVE} data are 
corrected for the interstellar neutral hydrogen column
assuming a neutral hydrogen column density 
of N(\ion{H}{1}) = $5\times 10^{17}{\rm\ cm^{-2}}$.} 
\end{figure}

\section{Comparison of Models with Data}

We compare here synthetic spectra generated from our model atmospheres with
the observed spectrophotometry.
The {\it EUVE} data are from Cassinelli et al. (1995).  The archival UV data
are from the S2/68 Ultraviolet Sky Survey Telescope (UVSST) on 
{\it TD-1} (\cite{td1}), the {\it Orbiting Astronomical Observatory 2} ({\it
OAO-2}~) (\cite{oao2}), the {\it International Ultraviolet Explorer} 
({\it IUE}~), and the {\it Goddard High Resolution Spectrograph} ({\it GHRS}~).
Optical spectrophotometric data are from \cite{13color} and
\cite{burnashev}.  The {\it TD-1} and {\it OAO-2} data were obtained
from the on-line archival catalogs of the Astronomical Data Center (ADC) at
Goddard Space Flight Center (GSFC).  The {\it IUE} data were obtained
from the National Space Science Data Center at GSFC.  With the
exception of the long wavelength primary (LWP) camera high-dispersion
spectra, the {\it IUE} data are all processed with the NEWSIPS
software (\cite{newsips}).  The {\it GHRS} data (\cite{heap95}) were
obtained from the {\it GHRS} data reduction facility at GSFC and
reduced using the CALHRS software in IDL maintained by the {\it GHRS}
Science Team.  Infrared J and K band fluxes are from \cite{jkdata} and
the {\it IRAS} color-corrected fluxes at 12 \micron\ and 25 \micron\
are from Cassinelli et al. (1995).  A synthetic spectrum from model N1
is compared with the observed spectrum of \epscma\ from 350 \AA\ to 25
\micron\ in Figure 1.  The stellar parameters of model N1 are listed
in Table 1.

%%
%% FIGURE 2
%%

\begin{figure}
\caption{The {\it EUVE} LW spectrum of \epscma\ from 350 \AA\ to
700 \AA\ compared with the 21750 K NLTE models (N1, N2, and N7) and a
ATLAS9 21000 K model.  The {\it EUVE} data are corrected for the
neutral interstellar column adopting a neutral hydrogen column density
of N(\ion{H}{1}) = $5\times 10^{17}{\rm\ cm^{-2}}$. The \phoenix\
model synthetic spectra have a resolution of 1 \AA.}
\end{figure}

%%
%% FIGURE 3
\begin{figure}
\caption{Model N2 in the EUV region with and without metal-line blanketing.
The synthetic spectra have a resolution of 1 \AA.} 
\end{figure}

\subsection{EUV Data}

Synthetic spectra from models N1 and N2 are compared with the
observed EUV spectrum of \epscma\ in Figure 2.  A plane-parallel LTE
ATLAS9 model (\Teff = 21000\,K, \Logg = 3.2, $\xi$ = 2 \kms) 
(\cite{kurucz92}) is shown for comparison.  
The effective temperature of models N1 and N2, \Teff = 21750 K, is at
the limit of the range established by Code {et~al.} (1976):
\Teff = 20990$\pm$760K.  The model surface gravity, \Logg = 3.2, 
and angular diameter, \AD = 0.77 mas, are within the established
limits: \Logg = 3.20$\pm$0.15, $\theta_d$ = 0.80$\pm$0.05 mas
(Cassinelli {et~al.} 1995).  Comparison of the position
of \epscma\ on the HR diagram with evolutionary tracks 
suggests the surface gravity is closer to \Logg = 3.5 (see section 4).
The synthetic spectra are scaled to match
the observed optical and near-IR spectrophotometry by 
adopting an angular diameter of 0.77 mas.  The ATLAS9 model is scaled
in the same way.  We emphasize here that the data have not been
normalized {\it -- all comparisons are made based on absolute fluxes.}

The {\it EUVE} absolute calibration uncertainty, which awaits a
complete in-orbit analysis, is estimated to be 25\%, while the
relative error is better than 10\% (\cite{vallerga}).  We correct the
{\it EUVE} data for transmission through the neutral interstellar
column based on cross sections of \cite{rumph94} and \cite{mormac83},
using the on-line ISM Transmission Tool available at the Center for
Extreme Ultraviolet Astronomy Web Page.  We adopt a neutral hydrogen
interstellar column density of N(\ion{H}{1}) = $5\times10^{17} $cm$^{-2}$, the
upper limit established by Gry et al. (1995).

Our synthetic spectra of models N1 and N2 very closely reproduce the H
I Lyman continuum observed by {\it EUVE}.  Model N2, which considers
ions of H, He, and CNO in NLTE, provides the best fit to the data
below the 504 \AA\ He I edge. However, the more physically realistic
model N1, which includes Mg II, Ca II, and Fe II-III in NLTE, predicts
less flux in the He I continuum.  At 400 \AA\ the N1 synthetic
spectrum is a factor of 3 below the observed flux.

A synthetic spectrum with the identical temperature structure of model
N1, but without metal line blanketing, shows that the shape and flux 
level of the \ion{H}{1} and \ion{He}{1} Lyman continua are not severely 
distorted by metal line blanketing
(see Figure 3).  At 1 \AA\ resolution, comparable to that of the {\it
EUVE} spectrum, the level of the apparent continuum is lowered by at
most 20\% between 500 \AA\ and 700 \AA.

%%
%% FIGURE 4

\begin{figure}
\caption{Temperature structures (left) and synthetic spectra
in the EUV (right) for models N1, N2 and N7.  A spherical
LTE model (L1) and an ATLAS9 spectrum are included for comparison.
The \phoenix\ synthetic spectra have a resolution of 1 \AA.}
\end{figure}

The temperature structures and EUV synthetic spectra of models N1, 
N2, and N7 are compared with an LTE model, L1, in Figure 4.  The
\ion{H}{1} Lyman continuum in both the NLTE and LTE models forms
at column masses greater than $10^{-4}$ g cm$^{-2}$ on the
plateau portion of the temperature structure common to these four
models at approximately 17200 K.  The overpopulation of the 
\ion{H}{1} ground-state in the NLTE models relative to the LTE model
causes the NLTE models to have a lower EUV flux
immediately below 912 \AA\ due to the larger 
cross section for \ion{H}{1} ionization from the ground-state. 
Far from the 912 \AA\ edge, just above the 504 \AA\ He I edge, where
the \ion{H}{1} ionization cross section approximately an order of
magnitude smaller, the NLTE and LTE fluxes are in better agreement.
NLTE appears not affect the temperature structure in the
\ion{H}{1} Lyman continuum forming layers. The 
\ion{H}{1} departure coefficients, however, regulate the flux in the 
Lyman continuum.  The NLTE models are significantly cooler at column
masses less than $10^{-5}$ g cm$^{-2}$ compared with the LTE model such
that the \ion{He}{1} edge, which is absent in the LTE model spectrum,
is quite strong in the NLTE models.  The larger relative fraction of
\ion{He}{1} and the overpopulation of the \ion{He}{1} ground-state in
the NLTE models results in a significantly lower EUV flux below 504
\AA\ compared with the LTE model.  An ATLAS9 model spectrum with
parameters \Teff = 21750 K and \Logg = 3.2 is shown in Figure 4 for
comparison.  Both the LTE and NLTE spherical \phoenix\ models predict a
higher EUV fluxes than the ATLAS model.  This is primarily due to the
significant differences in outer temperature structures between
spherical and plane-parallel models.  We discuss these differences in
detail in section 4.

%% FIGURE 5
\begin{figure}
\caption{Comparison of UV and optical spectrophotometry to 
model N1. With the exception of the {\it GHRS} LSA fluxes, 
the observed UV spectra lie below the model spectrum by
as much as 30\%.  The correction of Bohlin \& Holm (1984) 
has not been applied to the {\it OAO-2} data shown here.
The synthetic spectrum has a resolution of 5 \AA.}
\end{figure}

%%
%% FIGURE 6
%%

\begin{figure}
\caption{IUE high-dispersion data compared with model N1.  The
data have been scaled by a factor of 1.36 to match the model flux.  The
{\it IUE} data have been binned to 1 \AA\ to match the synthetic spectrum
resolution.}
\end{figure}
\subsection{Ultraviolet Data}

The extremely low column density, 
N(\ion{H}{1}) $\leq 5\times 10^{17}$ cm$^{-2}$
(\cite{gry95}), toward \epscma\ produces negligible dust extinction
in the ultraviolet and visible portions of the spectrum.
Consequently, no such corrections were applied to these data.

The pre-COSTAR {\it GHRS} large science aperture (LSA) ECH-B data
provide a measurement of the absolute UV continuum of \epscma\ at four
15 \AA\ wide orders between 2350 \AA\ and 2850 \AA.  Figure 5 shows
the comparison between a synthetic spectrum from model N1 and the UV
and optical spectrophotometry.  We plot the data and synthetic
spectrum in Figure 5 as $\lambda$ versus $\lambda^3 F_{\lambda}$ to
accentuate small differences in the slope and absolute levels of these
spectral energy distributions in the ultraviolet region.  The absolute
UV flux measurements from the {\it GHRS} ECH-B spectra are in good
agreement with model N1.  The {\it GHRS} spectra and model N1 show
$\sim$30\% more flux than measured by either {\it IUE}, {\it OAO-2},
or {\it TD-1}. The difference exceeds the published errors in the
absolute calibration of these data.  The published uncertainty for
{\it TD-1} is $\pm 20\%$ and the uncertainties for the {\it OAO-2} data 
are 30\% below 1300 \AA, 10\% between 1300 \AA\ and 1800 \AA, and 5\%
longward of 1800 \AA\ (\cite{td1oaoerr}).  The established uncertainty in
the absolute calibration for {\it IUE} is $\pm 10\%$ (\cite{bohlin86}).

In light of the agreement between model N1 and the {\it GHRS},
optical, and IR fluxes for \epscma, the {\it IUE}, {\it TD-1}, and
\oaotwo data appear to have a {\it wavelength independent} systematic
calibration error of $\sim$30\%.  
The \oaotwo data (\cite{oao2})
do not include correction of Bohlin \& Holm (1984) based on
the choice of \etauma\ (HR 5191, HD 120315; \Teff = 17000 K, \Logg =
4.25) as a fundamental UV standard star.  The UV spectral energy
distribution of \etauma, based on multiple {\it IUE} observations,
when compared with the \oaotwo data shows that the \oaotwo flux is
systematically larger between 1280 \AA\ and 2040 \AA, peaking at
$\sim$ 30\% larger around 1550 \AA. \cite{boholm} correct for this
\oaotwo flux excess.  Applying this same correction 
to the \oaotwo data for \epscma\ improves the agreement between these
data and the \iue and \tdone data sets.  

Only a single scan of \epscma\ was made with Spectrometer 1 (range:
1850 \AA\ - 3600 \AA) by \oaotwo (\cite{oao2}).  Two scans were
obtained with Spectrometer 2 (range: 1160 \AA\ - 1850 \AA).  Spectra
of stars for which we find good agreement between our model absolute
flux and the \oaotwo absolute flux, such as
\spica, consist of six scans in Spectrometer 1 and
three scans in Spectrometer 2.  Since the \oaotwo absolute flux of
\epscma\ longward of 1850 \AA\ is based on a single scan it may be
subject to greater error.  In addition, the \oaotwo
integrated absolute flux for \epscma\ from 1810 \AA\ to 3300 \AA\ as
listed by Code {et~al.} (1976) is 20\% greater than the value we compute
from the Code \& Meade (1979) data set.  For the integrated flux
of \spica\ in the same wavelength region 
we find no such disagreement between the two data sets.  
So, the Code {et~al.} fluxes
which are used to compute the effective temperature are in better
agreement with the {\it GHRS} absolute fluxes than the \oaotwo data shown
in Figure 5.  The absolute flux of the spectrophotometric standard
$\mu$ Col (HR 1996, HD 38666) at 2740 \AA\ as measured by {\it GHRS}
(3.1$\times 10^{-10}$ erg cm$^{-2}$ s$^{-1}$ \AA$^{-1}$; Heap {et~al.} 1995
and references therein) 
and \oaotwo (2.83$\times 10^{-10}$ erg cm$^{-2}$ s$^{-1}$ \AA$^{-1}$; Code
\& Meade 1979) differ by just less than 10\%. Like \epscma, a
single scan of $\mu$ Col was obtained by Spectrometer 1.

A discrepancy in the relative flux distribution was found
between model atmospheres and the {\it OAO-2} data for \betcma\
(\cite{betacma95}).  Applying the correction of \cite{boholm} to the
\oaotwo data of \betcma\ appears to resolve the discrepancy in 
the {\it relative} flux distribution for this star.  In this case,
however, neither \ghrs nor \iue large aperture data longward of 2000
\AA\ are available with which to check the {\it absolute} calibration.  
As with \epscma, the only a single scan of \betcma\ was obtained 
by Spectrometer 1.  We find the Code {et~al.} (1976) absolute integrated
flux between 1850 \AA\ and 3300\AA\ to be 10\% larger than the
same measurement from the Code \& Meade (1979) data set.
The absolute flux from the \iue large aperture SWP data for \betcma\
is in agreement with our model atmosphere for this star 
(Aufdenberg et al. 1997) \nocite{betacma}.

Here we investigate possible reasons for the discrepancy in the
absolute flux level between the model spectra and the \iue
observations.  Low-dispersion, large aperture, {\it IUE} observations
of a star as bright as \epscma\ (V = 1.5) required a rapid trailed
exposure.  Only two such exposures of \epscma\ exist, SWP54338 and
LWP30419, with exposure times of 0.184 s and 0.202 s respectively.
Rapid trails with exposure times less than a second, are generally not
as reliable as slower trailed spectra (\cite{Imhoff}).  For example,
the star may not have crossed the {\it IUE} large aperture along the
nominal path, thus affecting the overall flux levels.  We inspected
the low dispersion re-sampled {\it IUE} images (SILO) images of
\epscma\ and compared them to a rapid trailed SILO image of $\alpha$
Vir (SWP39248, 0.179 s), for which there is very good agreement
(within 5\%) between the measured absolute flux and our model
atmosphere (Aufdenberg et al., 1997)\nocite{betacma}.  The
\epscma\ SWP and LWP spectra are noticeably narrower than the width of
the $\alpha$ Vir spectrum perpendicular to the dispersion axis.  This
may account for the 30\% discrepancy in the {\it IUE} data.  An
additional source of error in the absolute calibration is the
extrapolation of the SWP and LWP camera sensitivity degradation
functions outside beyond valid date range.  Correcting for this
increases the absolute flux by $\sim$4\% (Imhoff 1997).

\subsubsection{{\it IUE} High Dispersion Data} 

The {\it IUE} low dispersion spectra were used by NEWSIPS to establish the
absolute calibration applied to the large aperture high resolution
spectra.  This explains why the {\it IUE} echelle spectra also
disagree with the model and the {\it GHRS} fluxes.  Despite the
absolute flux discrepancy between the {\it IUE} data and the model,
there is good agreement with the observed {\it relative} flux
distribution.  A comparison of model N1 and {\it IUE} high dispersion
spectra SWP54337 and LWP30183 is shown in Figure 6.  These high
dispersion images were chosen for their well exposed and non-saturated
continua.  Non-saturated regions of the other high dispersion spectra
we have examined show the same absolute flux level.

%%
%% FIGURE 7-9
%%
\begin{figure}
\caption{{\it GHRS} ECH-B observation Z14Z050MT compared to high resolution
synthetic spectrum of model N2}
\end{figure}

\begin{figure}
\caption{{\it GHRS} ECH-B observation Z14Z050XT compared to high resolution
synthetic spectrum of model N2.}
\end{figure}

\begin{figure}
\caption{{\it GHRS} ECH-B observation Z14Z051BT compared to high resolution
synthetic spectrum of model N2.}
\end{figure}

\subsubsection{{\it GHRS} Data}

The model N1 flux agrees with the {\it GHRS} continuum flux to within
$\pm 2\%$.  In addition to the agreement between the model and the
absolute flux level of the four {\it GHRS} ECH-B orders, high
resolution synthetic spectra of model N1 are in  reasonable agreement
with the observed line strengths and profiles (see Fiugres 7-9).  The
high resolution model spectra have been scaled by a small factor ($<
\pm$ 2\%) relative to the overall fit for the continuous energy
distribution to match the observed continuum level.  The widths
of the observed line profiles are slightly increased
as a result of the spherical aberration in these pre-COSTAR 
observations (\cite{heap95}).  The synthetic spectra were
rotationally broadened to 35 \kms\ (\cite{epscma95}) and corrected for
the heliocentric radial velocity of
\epscma, +27.4 \kms (\cite{gry95}).  No attempt was made to improve the 
match between the observed and synthetic spectra by changing
the chemical abundances or other model parameters.

\subsection{Infrared Data}

Comparison of model N1 mid-IR fluxes with the color-corrected {\it
IRAS} fluxes at 12 \micron\ and 25 \micron\ from Cassinelli et
al. (1995) indicates the presence of a slight mid-IR excess.  The IRAS
fluxes are 7\% larger at 12 \micron\ and 9\% larger at 25 \micron\
compared with model N1.  Model N1 fluxes are within 1.1 $\sigma$ and 
1.2 $\sigma$ of the mean IRAS fluxes at 12 \micron\ and 25 \micron,
respectively.  The model predicts 9\% and 10\% more flux at 12 \micron\
and 25 \micron\ respectively, compared with the ATLAS9 \Teff = 21000 K,
\Logg = 3.2, $\theta_d$ = 0.80 mas model.

%%%
%%%FIGURE 10
%%%
\begin{figure}
\caption{Error box for \epscma\ based on the observational uncertainties
plotted along with 
evolutionary tracks (Schaller {et~al.} 1992) of solar metallicity
on a theoretical HR diagram.  The error bars reflect the combined
1 $\sigma$ errors in the measured total flux and the measured angular diameter
(Code {et~al.} 1976), and the standard error in the {\it HIPPARCOS}
parallax. The diamond symbol represents the position of \epscma\ from
the best fit values based on the comparison of the 
synthetic spectra to the spectrophotometry adopting the mean 
{\it HIPPARCOS} distance.}
\end{figure} 

\section{Discussion and Conclusions}
\subsection{Stellar Parameters and the Local ISM}

We find that the observed flux distribution between 350 \AA\ and 25
\micron\ can be reasonably well represented by our spherical,
hydrostatic, non-LTE,
line blanketed model atmosphere by adopting model parameters within
the range of uncertainty of the established stellar parameters.
Adopting the {\it HIPPARCOS} distance of 132$^{+11}_{-9}$ pc and the
best fit model parameters, $\theta_d$=0.77 and \Logg = 3.2, yield a
radius of 11.3$\pm$ 1.1 $R_{\sun}$ and a mass of 7.0$^{+1.2}_{-0.9}$
$M_{\sun}$ for \epscma.  The uncertainties in radius and mass reflect
only the uncertainty in the {\it HIPPARCOS} parallax.  Our value for
the radius of \epscma\ is $\sim$ 30\% smaller than that quoted by
Cassinelli { et~al.} (1995) primarily due to the {\it HIPPARCOS}
distance, which puts \epscma\ 30\% closer than previously
believed. This smaller radius reduces the mass by 54\% for the same
surface gravity.  The surface gravity of \Logg = 3.2 appears to be too
low by approximately 0.3 dex.  Based on a comparison with evolutionary
tracks (see Figure 10), \epscma\ has a mass of nearly 12$M_{\sun}$,
however, the measured radius with a gravity of \Logg = 3.2$\pm$0.15
yields a mass of 7.4$\pm$3.0$M_{\sun}$.  An atmosphere model (model N7)
computed with \Logg = 3.5, but with otherwise identical stellar
parameters as model N1, does not show a significantly different EUV
flux distribution (see Figure 2).  A possible resolution to this mass
discrepancy may result from a reanalysis of the Balmer line fits used to
establish the surface gravity.  Apparently the only published high
dispersion Balmer profiles are photographic data from Buscombe (1970).
We leave this analysis for future work.

The predicted number of hydrogen and helium ionizing photons at the
stellar surface from model N1 are $1.59\times10^{21}$ photons
cm$^{-2}$ s$^{-1}$ (0 \AA\ - 912 \AA) and $1.13\times10^{18}$ photons
cm$^{-2}$ s$^{-1}$ ( 0 \AA\ - 504 \AA) respectively.  The best fitting
model EUV flux (model N2) at the stellar surface is listed in Table 2.  Using
the best fit angular diameter, 0.77 mas, the calculated unattenuated
hydrogen ionizing flux (0 \AA\ - 912 \AA) and neutral hydrogen
photoionization rate at the Local Cloud surface are 2290 photons
cm$^{-2}$ s$^{-1}$ and $1.06\times10^{-14}$ s$^{-1}$ respectively.
These values are consistent with those of Vallerga \& Welsh (1996) who
compute values more than twice as large but adopt a correspondingly
larger neutral hydrogen column density of $9\times10^{17} {\rm\
cm^{-2}}$ and a distance of 188 pc.

%FIGURE 11 - temp plot
\begin{figure}
\caption{Temperature structures 
for a variety of spherical and plane-parallel models listed in Table
1. An ATLAS9 structure is included for comparison. 
Spherical
line blanketed models show significantly warmer temperature structures
compared with plane-parallel line blanketed models.  NLTE models with
no metal line blanketing show identical structures for the two
geometrical treatments.}
\end{figure}

%Figure 12 - spec plot
\begin{figure}
\caption{Synthetic spectra in the EUV
for a variety of spherical and plane-parallel models listed in Table
1. An ATLAS9 spectrum is included for comparison.  Spherical
line blanketed models predict significantly larger EUV fluxes compared
with plane-parallel line blanketed models.  NLTE line blanketed models
predict similar or lower EUV fluxes than LTE models due to the
overpopulation \ion{H}{1} and \ion{He}{1} ground states in the NLTE
treatment.  NLTE models with no metal line blanketing show identical
EUV spectra for the two geometrical treatments.}
\end{figure}

\subsection{Effects of Sphericity and Line Blanketing on the EUV
Continuum}

In this section we explore the sensitivity of our model atmospheres to
the treatment of the problem with spherical or plane-parallel
geometry, the degree of line blanketing, and other model parameters.
In Figure 4 we show synthetic spectra from our LTE and NLTE
spherical line blanketed models compared with a LTE plane-parallel
line blanketed model from Kurucz with same effective temperature and
surface gravity.  We find that the differences in the predicted EUV flux
between our models and that of Kurucz are primarily 
due to the differences in the
temperature structures between line blanketed spherical models and line
blanketed plane-parallel models.  To explore these differences, we
computed a grid of models, all of which have an effective temperature
of \Teff = 20990 K, a surface gravity of \Logg = 3.2, and a turbulent
velocity of $\xi$ = 2 km s$^{-1}$.  These models differ in the geometry of
the calculation (spherical or plane-parallel), the use and degree of
LTE or NLTE line blanketing, the presence or absence of LTE line
scattering, the value of the outer boundary pressure, the value of the
outer boundary continuum optical depth, and the resolution of the
wavelength grid in the model calculation.  These models are listed in
Table 1.

We find that the combination of parameters in our models which has the
most significant effect on the temperature structure, and therefore
the EUV  flux level, is the inclusion of both metal line blanketing and
a spherical geometry.  The temperature structure, in particular
temperatures in the region responsible for the EUV continuum, is only
slightly affected, if at all, by different hydrostatic outer
boundary conditions, the degree of line blanketing, the presence of
LTE line scattering, or the resolution of the wavelength grid.  We now
discuss briefly each of these test models.

\subsubsection{Sphericity and Line Blanketing} 

In Figure 11 we compare the temperature structures of six LTE and NLTE
models with and without line blanketing.  The ATLAS9 temperature
structure is shown for comparison.  The temperature structures of the
line blanketed models fall into two groups. The spherical models
show a plateau structure near 16600 K between column masses
1 g cm$^{-2}$ and $10^{-3}$ g cm$^{-2}$, while the plane-parallel
models show a plateau structure near 13300 K between 
column masses $10^{-2}$ g cm$^{-2}$ and  $10^{-4}$ g cm$^{-2}$.  The
NLTE and LTE models have nearly the same temperature structure in
these plateau regions and differ substantially only beyond the
plateau at lower mass columns.  The differences in the temperature
structures between the two geometries have significant effects on the
flux levels of the synthetic EUV continua as shown in Figure 12.  For
example, the EUV flux at 750 \AA\ is a factor of 6 higher
for the spherical LTE line blanketed model (L2) than the
plane-parallel LTE line blanketed model (L3).   

The \ion{H}{1} Lyman continuum of the LTE spherical model (L2) forms at
16500 K, while for the LTE plane-parallel model (L3) the Lyman continuum
forms at approximately 14500 K.  The Lyman continua of the NLTE
line blanketed models form at the same column masses in the atmosphere as the
LTE models, but, the overpopulation of the ground states in the
NLTE models further affects the flux levels of the \ion{H}{1} and
\ion{He}{1} Lyman continua.  Our LTE, line blanketed, plane-parallel
synthetic spectrum predicts the same flux in the \ion{H}{1} Lyman
continuum as the ATLAS9 spectrum, but these models predict different
flux levels for the \ion{He}{1} continuum below 504 \AA.  This is due
to the different temperature structures predicted by these models for
column masses less than $10^{-2}$ g cm$^{-2}$.  We do not yet
understand the reasons for the differences between these models in
this region.  Differences in the two codes may exist in the treatment
of metal line blanketing.  In addition, important differences may
exist between the two codes for the equation-of-state, which is
computed by \phoenix\ in a completely independent way from that of
ATLAS, e.g., by using different energy level data.

In contrast to the large differences in the temperature structures and
EUV continua for the line blanketed models with respect to a spherical
versus a plane-parallel geometry, the NLTE H-He only models (N5 and
N6), with no metal line blanketing, show nearly identical temperature
structures and EUV spectra for both geometries.  Therefore, the
significant differences in the model EUV spectra of early B stars for a
particular treatment of the geometry is the result of the inclusion of
metal line blanketing.  The Lyman continuum of the unblanketed pure
H-He NLTE model forms at 13500 K and the flux level at 750 \AA\ is
more than a factor of 15 lower than that predicted by the spherical
NLTE model (N3) with line blanketing.

%%%%%%%%%%%%%%%%%%%REVISED PHYSICAL DESCRIPTION%%%%%%%%%%%%%%%%%%%%%%%%

We suggest that this can be explained physically as follows.
Including metal line blanketing in a model atmosphere produces the
well known effects of backwarming and surface cooling relative to the
temperature structure in an unblanketed atmosphere.  These effects are
seen by comparing the temperature structures of plane-parallel model,
L3 (metal line blanketed) with model N6 (H and He opacity only) in
Figure 11.  The surface layers, mass depths $<$ 10$^{-5}$ g cm$^{-2}$,
of the L3 structure are cooler relative to N6, while the deeper
layers, depths $>$ 10$^{-5}$ g cm$^{-2}$, are warmer relative to N6
until depths greater than 1 g cm$^{-2}$ are reached.  In the layers of
the atmosphere sensitive to blanketing, the metal lines produce an
opacity, primarily in the ultraviolet, that heats the layers in
radiative equilibrium immediately below and cools the layers
immediately above, producing a steeper temperature gradient relative
to the model lacking this metal opacity.

These backwarming and surface cooling effects of line blanketing are
seen in the spherical models, but with a difference.  Looking at the
temperature structures of models L2 (spherical) and L3 (PP), these
models have basically the same slope from 10$^{-7}$ g cm$^{-2}$ up to
their respective plateaus.  The spherical and PP structure shapes are
very similar from the plateaus outward.  However, the PP model
temperature structure flattens out at 10$^{-5} $g cm$^{-2}$, while the
spherical model temperature slope is sustained into deeper layers (to
10$^{-4}$ g cm$^{-2}$) before flattening out to its plateau.  The
spherical temperature structure maintains the same slope, from
10$^{-7}$ g cm$^{-2}$ inward to the plateau, as the PP case, but to
deeper layers.  In other words, the temperature gradient outward from
the plateau begins at greater depth in the spherical model compared
with the PP model.  Effectively, in the spherical case, the surface
cooling begins ``earlier'', moving outward in the structure.
Therefore in the spherical case, the resultant backwarmed layers
(depths $>$ 10$^{-4}$ g cm$^{-2}$) which form the plateau, are at a
higher temperature beneath the cooling layers (depths $<$ 10$^{-4}$ g
cm$^{-2}$) relative to the plateau in the PP model, where cooling
begins further outward.  Since the surface cooling begins at a greater
depth in the spherical case, the backwarming is shifted to deeper
layers relative to the PP case.  These deeper layers are intrinsically
warmer, so the temperature of the spherical model backwarmed plateau
sits at a higher temperature than the PP model plateau.

One important factor which shifts the surface cooling to greater
depths in the spherical models is the reduced optical depth seen by
photons in a spherical geometry relative to that in a semi-infinite
plane-parallel geometry at a given physical depth.  Immediately below
the cooling layers, the enhanced photon escape probability increases
the populations in the lower levels of the metal transitions, thereby
increasing the local opacity.  In response, the layers immediately
below must heat up in order to transport the same flux in the presence
of the increased line opacity and reduced frequency bandwidth.  As a
result, in the spherical models, backwarming begins at a greater depth
than in the PP models.  

The result that the H-He only models (N5 and N6) are not sensitive to
the geometry is likely because the H-He line opacity, which is nearly
absent in the ultraviolet, has a much weaker effect on the temperature
structure relative to the metal lines.  The enhanced escape
probability for photons in the spherical geometry of a H-He atmosphere
has a much weaker effect on the cooling rates and therefore the
temperature structure.

Our results can be explained on the basis of the geometrical effects
in combination with the metal line blanketing in LTE.  NLTE processes,
which are necessary for the detailed modeling of the observed spectra
(for instance, the presence of the \ion{He}{1} 504 \AA\ edge), 
clearly have effects on the temperature
structure, but these are less significant that those effects due to
the combination of a spherical geometry and LTE metal-line blanketing
alone.

\subsubsection{Number of LTE Blanketing Lines} 

%Figure 13 - LTE lines plot
\begin{figure}
\caption{Temperature structures (left) and synthetic spectra
in the EUV (right) for models which consider different
numbers blanketing lines in LTE.  The temperature structure
and EUV spectrum are sensitive to approximately the strongest 10,000 
lines and a largely unaffected by millions of weaker metal lines.}
\end{figure}

We have tested the sensitivity of the temperature structure to the
number of blanketing atomic lines.  For the typical LTE model the
number of lines used in the calculation is approximately
$2\times10^6$. The temperature structure, however, is mainly
determined by approximately the 50000 strongest lines.  Figure 13
compares the temperature structures and synthetic spectra in the EUV
for a grid of six spherical models (L2, L11-L15) each with a different
number of atomic blanketing lines from 2095181 lines to only the
strongest 217 lines.

The level of the plateau in the temperature structure, between column
masses $10^{-1}$ g cm$^{-2}$ and $10^{-3}$ g cm$^{-2}$, is 600 K lower
for 217 lines compared with 2095181 lines.  The level of the EUV
continuum at 750 \AA\ is 30\% lower for 217 lines compared with
2095181 lines.  In contrast, a line blanketed plane-parallel model
(L3) is 2800 K cooler than an otherwise similar line blanketed
spherical model and their predicted EUV fluxes differ at 750 \AA\ by
more than a factor of 6.  We conclude the EUV continuum flux in LTE is
regulated by the several thousand strongest blanketing lines and that
there is little affect on this EUV continuum from the inclusion of
millions of weak lines used in the model.

%Figure 14 - scattering plot
\begin{figure}
\caption{Temperature structures (left) and synthetic spectra
in the EUV (right) for models which have either a large value, (95\%), for
the LTE line scattering albedo, or no LTE line scattering.  The
temperature structure is generally slightly cooler with the inclusion
of line scattering and the EUV continuum flux is slightly weaker.}
\end{figure}

\subsubsection{LTE Line Scattering} 

In order to test the effect of LTE line scattering on the temperature
structure of LTE line blanketed models, we have computed both spherical
and plane-parallel models (L2-L5) with values for the depth independent
line scattering albedo of $\alpha$ = 95\% and $\alpha$ = 0\%.
Neglecting line scattering corresponds to pure LTE.  The temperature
structures and EUV synthetic spectra are compared in Figure 14.
Including LTE line scattering reduces the temperature down to
column masses of 1 g cm$^{-2}$ and consequently also reduces the flux
level of the EUV continuum.  The line scattering in our LTE models
changes the level of the EUV continuum by factors of $\sim$1.5 in the
EUV continuum at 750 \AA\ and $\sim$2.5 at 450 \AA.  These effects are
less significant than those due to a plane-parallel geometry.

%%
%% FIGURE 15 - Pressure Test
%%

\begin{figure}
\caption{Temperature structures (left) and synthetic spectra
in the EUV (right) for models with different values of the
outer boundary gas pressure (in dynes cm$^{-2}$). 
The temperature structure and EUV flux are little affected 
by the value of the outer pressure.}
\end{figure}

%%
%% FIGURE 16 - Tau Test
%%
\begin{figure}
\caption{Temperature structures (left) and synthetic spectra
in the EUV (right) for models with different values of the
outer boundary continuum optical depth. The temperature structure and
EUV flux are not affected by the value of the boundary optical depth,
however, they are significantly affected by the geometric
treatment of the model atmosphere.}
\end{figure}

\subsubsection{Outer Boundary Conditions}   

Our standard value for the outer boundary gas pressure, $P_{\rm gas} =
10^{-4}$, is one which is traditional (Kurucz 1970).  To test the
effect of this assumption we have converged models (L2, L6, L7) for
outer gas pressures of $10^{-4}$, $10^{-5}$, and $10^{-6}$ dynes
cm$^{-2}$.  These models produce nearly identical temperature
structures and identical synthetic spectra (see Figure 15).  We have
also converged models (L9 and L10) for outer continuum optical depth
of $10^{-6}$ and $10^{-10}$.  Adjusting the outer boundary optical
depth from $10^{-10}$ to $10^{-6}$, similar to the ATLAS9 model,
appears to have no effect on the temperature structure or synthetic
spectra for a given geometry, spherical or plane-parallel (see Figure
16). Realistic outer boundary conditions for \epscma\ are probably not
hydrostatic.  We leave the investigation of the effects of more
realistic boundary conditions, such as a stellar wind, for future work.

%FIGURE 17 -- Lambda grid test
\begin{figure}
\caption{Temperature structures (left) and synthetic spectra
in the EUV (right) for models with different resolutions of
the wavelength grid used in the model computation. 
The temperature structure and the EUV flux are not significantly 
affected by the resolution of the wavelength grid.}
\end{figure}

\subsubsection{Wavelength Grid Resolution}

We have computed a small grid of spherical LTE line blanketed models
to test the sensitivity of the converged temperature structure to the
number of wavelength grid points employed in the model calculation.
Figure 17 shows the temperature structures and EUV synthetic spectra
for three models which were computed on wavelength grids of 11074,
22148, and 44296 points respectively.  The 11074 grid points model has
a slightly warmer temperature structure of column masses less than
$10^{-5}$ g cm$^{-2}$ and a more linear slope near $10^{-4}$ g
cm$^{-2}$.  At greater depths the temperature structures of the three
models are nearly identical and the synthetic EUV spectra of the three
models are also nearly identical. The NLTE models that were computed for
comparison with observational data use more than 60000 wavelength
points.

\subsection{Summary}

The continuous energy distribution of
\epscma\ from the EUV to the IR is reasonably well fit by 
a hydrostatic model atmosphere that neglects a stellar wind, magnetic
fields, and X-ray heating.  We find the flux level of the helium
continuum to be sensitive to the number of ions and transitions
treated in NLTE.  Our most realistic spherical, NLTE, hydrostatic
model is unable to reproduce the slope of the observed spectrum below
504 \AA.  The slope may be affected by a stellar wind and non-thermal
heating processes which may alter the ionization structure of the
atmosphere and the population of the He I ground state.  Our model
fit, based on absolute fluxes together with the {\it HIPPARCHOS}
distance, yields fundamental parameters for \epscma.

Our \Teff = 21000 K, \Logg = 3.2, spherical NLTE model predicts more
than twice as many hydrogen ionizing photons and over 200 times more
neutral helium ionizing photons compared to a standard hydrostatic
plane-parallel LTE model with the same stellar parameters.  This is
important since the EUV flux from \epscma\ may not be
peculiar, and is
well represented by our spherical hydrostatic model atmosphere.  It
touches directly on the contribution of early B stars 
to the diffuse UV radiation field and the photoionzation of the
interstellar medium.

We find significant differences in the strength of the predicted EUV
flux between our line blanketed spherical and line blanketed
plane-parallel models.  A more realistic model treatment of early B
giants with a spherical geometry and NLTE metal line blanketing
results in the prediction of significantly larger EUV fluxes compared
with plane-parallel models.  This result appears to explain a large
part of the reported discrepancy between the observed EUV flux of
early B giant \epscma\ and the EUV flux predicted by plane-parallel
LTE and NLTE line blanketed model atmospheres.

\acknowledgments

We thank the referee, I. Hubeny, for his careful reading and detailed
comments which significantly improved this paper, D.H. Cohen for
providing us with the reduced EUVE data and for useful comments on
early drafts, R. L. Kurucz for the ATLAS atmospheric structure,
C. Imhoff for details on the {\it IUE} absolute calibration and rapid
trailed spectra and S. Starrfield for useful comments on the draft.
This work made use of the SIMBAD database, Strasbourg, France.  JPA
acknowledges support from an ASU NASA Space Grant Fellowship.  This
work was supported in part by NASA ATP grant NAG 5-3018 and LTSA grant
NAG 5-3619 to the University of Georgia, by NASA LTSA grants NAGW 4510
and NAGW 2628, NASA ATP grant NAG 5-3067 and by NSF grant AST-9417057
to Arizona State University, and by NSF grant AST-9417242, NASA grant
NAG5-3505 and an IBM SUR grant to the University of Oklahoma.  Some of
the calculations presented in this paper were performed on the IBM SP2
of the UGA UCNS, at the Cornell Theory Center (CTC), the San Diego
Supercomputer Center (SDSC) and the NCSA, with support from the
National Science Foundation, and at the NERSC with support from the
DoE. We thank all these institutions for a generous allocation of
computer time.

\clearpage

%
%TABLE 1
%
\begin{table*}
\tablenum{1}
\caption{Models \label{tabmod}}
\begin{center}
\begin{tabular}{llcccccl}
Model &Type &\Teff (K) &$\log(g)$ &$\xi$ (\kms) &Geometry &LTE Scattering 
&Comment\\

\tableline

N1 &NLTE  &21750 &3.2 & 2.0 &Spherical      &yes &H-He-CNO-Ca-Mg-Fe\\
N2 &NLTE  &21750 &3.2 & 2.0 &Spherical      &yes &H-He-CNO  \\
N3 &NLTE  &20990 &3.2 & 2.0 &Spherical      &yes &H-He-CNO  \\
N4 &NLTE  &20990 &3.2 & 2.0 &Plane-Parallel &yes &H-He-CNO \\
N5 &NLTE  &20990 &3.2 & 2.0 &Spherical      &no  &H-He only\\
N6 &NLTE  &20990 &3.2 & 2.0 &Plane-Parallel &no  &H-He only \\
N7 &NLTE  &21750 &3.5 & 2.0 &Spherical      &yes &H-He-CNO-Ca-Mg-Fe\\

L1 &LTE   &21750 &3.2 & 2.0 &Spherical      &yes &  \\
L2 &LTE   &20990 &3.2 & 2.0 &Spherical      &yes &  \\
L3 &LTE   &20990 &3.2 & 2.0 &Plane-Parallel &yes &  \\
L4 &LTE   &20990 &3.2 & 2.0 &Spherical      &no  &  \\ 
L5 &LTE   &20990 &3.2 & 2.0 &Plane-Parallel &no  &  \\
L6 &LTE   &20990 &3.2 & 2.0 &Spherical      &yes & P$_{\rm gas} = 10^{-5}$ \\
L7 &LTE   &20990 &3.2 & 2.0 &Spherical      &yes & P$_{\rm gas} = 10^{-6}$\\
L8 &LTE   &20990 &3.2 & 2.0 &Spherical    &yes & $\tau_{\rm min} = 10^{-6}$ \\
L9 &LTE  &20990 &3.2 & 2.0 &Plane-Parallel &yes & $\tau_{\rm min} = 10^{-6}$\\ 
L10 &LTE  &20990 &3.2 & 2.0 &Spherical     &yes 
&$\kappa_{\rm lc}/\kappa_{\rm cont} > 1$  \\
L11 &LTE  &20990 &3.2 & 2.0 &Spherical     &yes 
&$\kappa_{\rm lc}/\kappa_{\rm cont} > 10$  \\
L12 &LTE  &20990 &3.2 & 2.0 &Spherical     &yes 
&$\kappa_{\rm lc}/\kappa_{\rm cont} > 10^2$  \\
L13 &LTE  &20990 &3.2 & 2.0 &Spherical     &yes 
&$\kappa_{\rm lc}/\kappa_{\rm cont} > 10^4$  \\
L14 &LTE  &20990 &3.2 & 2.0 &Spherical     &yes 
&$\kappa_{\rm lc}/\kappa_{\rm cont} > 10^5$  \\
L15 &LTE  &20990 &3.2 & 2.0 &Spherical     &yes 
& 22148 wavelength points \\
L16 &LTE  &20990 &3.2 & 2.0 &Spherical     &yes 
& 44296 wavelength points \\

\end{tabular}
\end{center}
\end{table*}

\clearpage

%
%TABLE 2
%

\begin{table*}
\tablenum{2}
\caption{Model N2 EUV Surface Flux \label{tabparm}}
\begin{center}
\begin{tabular}{cccc}
$\lambda$ (\AA) &Log F$_{\lambda}$\tablenotemark{a}
&$\lambda$ (\AA) &Log F$_{\lambda}$\tablenotemark{a}\\

 200  &   -3.511  & 590  &    7.376\\
 215  &   -2.432  & 605  &    7.448\\
 230  &   -0.428  & 620  &    7.557\\
 245  &    0.388  & 635  &    7.616\\
 260  &    0.984  & 650  &    7.584\\
 275  &    1.793  & 665  &    7.694\\
 290  &    2.328  & 680  &    7.841\\
 305  &    2.805  & 695  &    7.901\\
 320  &    3.072  & 710  &    7.847\\
 335  &    3.493  & 725  &    7.979\\
 350  &    3.843  & 740  &    8.054\\
 365  &    4.261  & 755  &    8.097\\
 380  &    4.535  & 770  &    8.137\\
 395  &    4.786  & 785  &    8.175\\
 410  &    5.040  & 800  &    8.212\\
 425  &    5.225  & 815  &    8.245\\
 440  &    5.456  & 830  &    8.280\\
 455  &    5.654  & 845  &    8.305\\
 470  &    5.827  & 860  &    8.334\\
 485  &    5.978  & 875  &    8.362\\
 500  &    6.138  & 890  &    8.223\\
 515  &    6.726  & 905  &    8.409\\
 530  &    6.807  & 920  &    8.990\\
 545  &    7.140  & 935  &    8.830\\
 560  &    7.115  & 950  &    8.877\\
 575  &    7.324  & 965  &   10.133\\

\end{tabular}
\end{center}
\tablenotetext{a}{Units: erg cm$^{-2}$ s$^{-1}$ \AA$^{-1}$ at the
stellar surface.}
\end{table*}

\end{document}